%% file: main.tex
\documentclass[9pt,sigconf,screen]{acmart}
\usepackage{color}
\usepackage{booktabs} 
\usepackage{xcolor}
\usepackage{multirow}
\usepackage{graphicx} 
\usepackage{ifthen}
\usepackage{academicons}

\newcommand{\paranl}[1]{{\bf #1: }}
\graphicspath{{figure/}{figures/}}
\setcopyright{none}

\copyrightyear{2023} 
\acmYear{2023} 
\acmConference[ACM ICN '23]{10th ACM Conference on Information-Centric Networking}{October 9--10, 2023}{Reykjavik, Iceland}
\acmBooktitle{10th ACM Conference on Information-Centric Networking (ACM ICN '23), October 9--10, 2023, Reykjavik, Iceland}

\begin{document}
\title{SoK: Distributed Computing in ICN}
\titlenote{The manuscript has been accepted to appear in the proceedings of ACM ICN 2023.}

\input{authors.tex}

\begin{abstract}
  \input{abstract}

\end{abstract}


\maketitle

\input{intro}

\input{discompicn.tex}

\input{enabler.tex}
\input{protocol.tex}
\input{orchestration.tex}

\input{application}

\input{conclusion}

\input{ack}

\bibliographystyle{acm}
\bibliography{icn,dku,wei} 
\end{document}

%% file: authors.tex

\newcommand{\ustgz}{The Hong Kong University of Science and Technology (Guangzhou)}
\newcommand{\aff}{
\affiliation{
  \institution{\ustgz}
  \city{Guangzhou} 
  \country{China}
}
}
\renewcommand{\ustgz}{HKUST(GZ)\dag}

\settopmatter{authorsperrow=6}

\author{Wei Geng}\orcid{0000-0001-5970-3550}
\aff

\author{Yulong Zhang}\orcid{0000-0003-0003-8626}
\aff

\author{Dirk Kutscher}\orcid{0000-0002-9021-9916}
\authornote{D. Kutscher is the corresponding author. dku@ust.hk\\
\dag HKUST(GZ) is The Hong Kong University of Science and Technology (Guangzhou).}
\aff

\author{Abhishek Kumar}\orcid{0000-0003-4383-7225}
\affiliation{
    \institution{University of Oulu}
    \city{Oulu}
    \country{Finland}
}

\author{Sasu Tarkoma}\orcid{0000-0003-4220-3650}
\affiliation{
    \institution{University of Helsinki and University of Oulu}
    \city{Helsinki}
    \country{Finland}
}

\author{Pan Hui}\orcid{0000-0001-6026-1083}
\affiliation{
    \institution{ \ustgz and University of Helsinki}
    \city{Guangzhou}
    \country{China}
}


%% file: abstract.tex
Information-Centric Networking (ICN), with its data-oriented operation and generally more powerful forwarding layer, provides an attractive platform for distributed computing. This paper provides a systematic overview and categorization of different distributed computing approaches in ICN encompassing fundamental design principles, frameworks and orchestration, protocols, enablers, and applications. We discuss current pain points in legacy distributed computing, attractive ICN features, and how different systems use them. This paper also provides a discussion of potential future work for distributed computing in ICN.

%% file: intro.tex
\section{Introduction}
\label{sec:intro}

Distributed computing -- a model where distributed components for
computing and storage communicate over a network to form a larger
system -- is the basis for all relevant applications on the
Internet. Based on well-established principles
\cite{lamport1990distributed}, different mechanisms, implementations,
and applications have been developed that form the foundation of the
modern Web.

The Internet with its stateless forwarding service and end-to-end
communication model \cite{moore2023ruined} promotes certain types of
communication for distributed computing. For example, IP addresses
and/or DNS names provide different means for identifying computing
components. Reliable transport protocols (e.g., TCP, QUIC) promote
interconnecting modules. Communication patterns such as REST
\cite{fielding00} and protocol implementations such as HTTP enable
certain types of distributed computing interactions, and security
frameworks such as TLS and the web PKI \cite{10066507} constrain the
use of public-key cryptography for different security functions.

With currently available Internet technologies, we can observe a
relatively succinct layering of networking and distributed computing,
i.e., distributed computing is typically implemented in overlays with
Content Distribution Networks (CDNs) being prominent and ubiquitous
example. Recently, there has been growing interest in revisiting this
relationship, for example by the IRTF Computing in the Network
Research Group (COINRG)\footnote{\url{https://irtf.org/coinrg}} --
motivated by advances in network and server platforms, e.g., through
the development of programmable data plane platforms
\cite{10.1145/3447868} and the development of different types of
distributed computing frameworks, e.g., stream processing
\cite{8864052} and microservice
frameworks. \cite{10.1145/3393822.3432339} This is also motivated by
the recent development of new distributed computing applications such
as distributed machine learning (ML) \cite{10.1145/3377454} and emerging
new applications such as Metaverse suggest new levels of scale in
terms of data volume for distributed computing and the pervasiveness
of distributed computing tasks in such systems. There are two research
questions that stem from these developments:


1) How can we build distributed computing systems in the network that
can leverage the on-path location of compute functions, e.g., optimally
aligning stream processing topologies with networked computing
platform topologies?

2) How can the {\em network} support distributed computing in general,
so that the design and operation of such systems can be simplified,
but also so that different optimizations can be achieved to improve
performance and robustness?

ICN (we focus mainly on CCNx/NDN-based ICN in this paper) with its
data-oriented operation and generally more powerful forwarding layer
provides an attractive platform for distributed computing. Several
different distributed computing protocols and systems have been
proposed for ICN, with different feature sets and different technical
approaches, including Remote Method Invocation (RMI) as an interaction
model as well as more comprehensive distributed computing
platforms. RMI systems such as RICE \cite{Krl2018RICERM} leverage the
fundamental named-based forwarding service in ICN systems
\cite{Wissingh2020InformationCentricN} and map requests to \textit{Interest}
messages and method names to content names (although the actual
implementation is more intricate as we will explain below). Method
parameters and results are also represented as content objects, which
provides an elegant platform for such interactions.

\begin{figure}[htbp]
    \centering
    \includegraphics[width=\columnwidth]{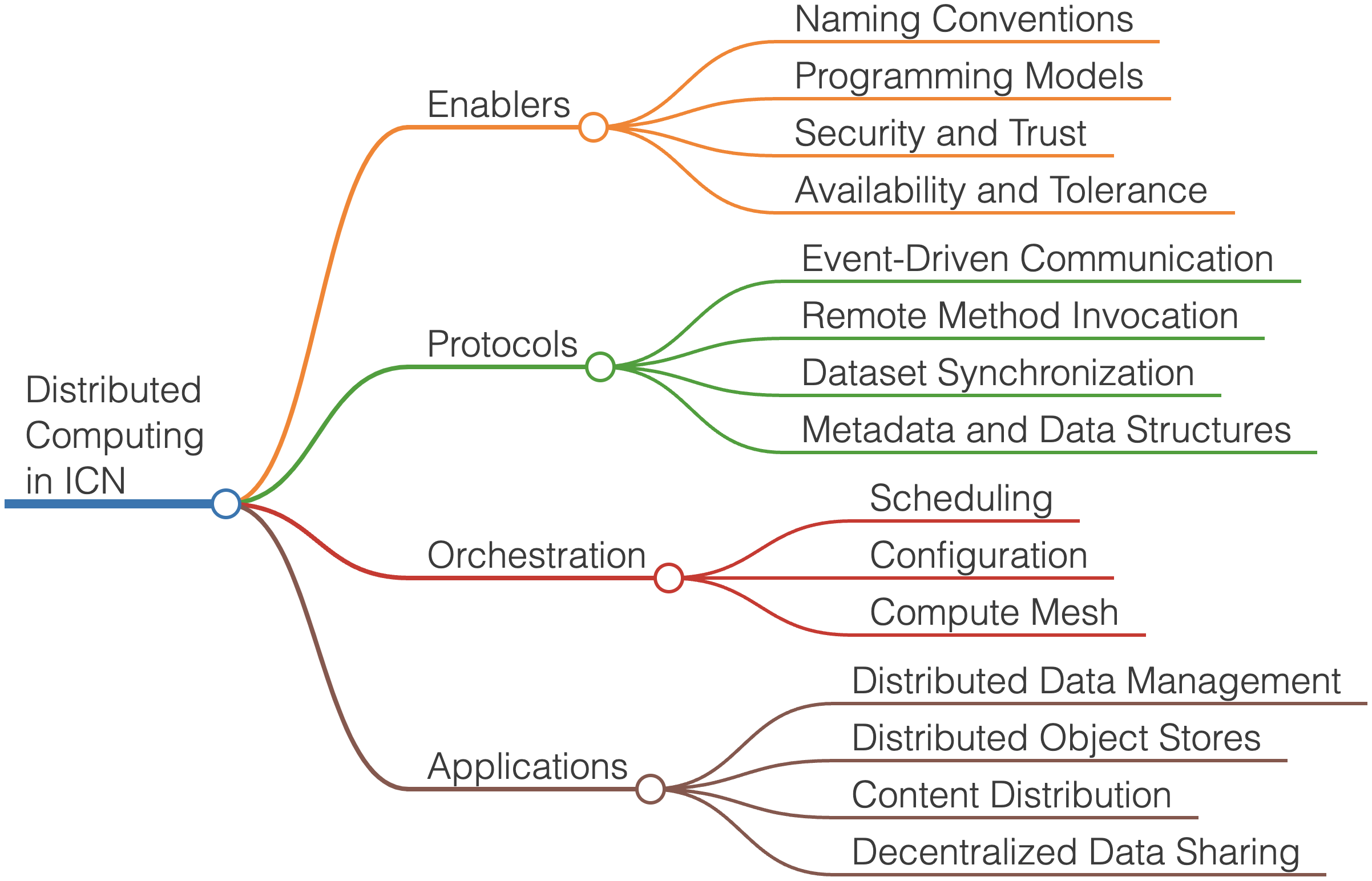}
    \caption{A category of distributed computing systems in ICN.}
    \label{fig:taxonomy}
\end{figure}

This SoK paper provides a comprehensive analysis and understanding of
distributed computing systems in ICN, based on a survey of more than
50 papers. Naturally, these different efforts cannot be directly
compared due to their difference in nature. We categorized different
ICN distributed computing systems, and individual approaches and
highlighted their specific properties.
The scope of this study is {\em technologies for \textbf{ICN-enabled
    distributed computing}}. Specifically, we divide the different
approaches into four categories, as shown in figure
\ref{fig:taxonomy}: enablers, protocols, orchestration, and
applications. The contributions of this study are as follows:

\begin{enumerate}
    \item A discussion of the benefits and challenges of distributed
      computing in ICN.
    \item A categorization of different proposed distributed computing
      systems in ICN.
    \item A discussion of lessons learned from these systems.
    \item A discussion of existing challenges and promising directions
      for future work.
\end{enumerate}
The rest of this paper is structured as follows
section \ref{sec:discompicn} describes relevant general distributed
computing concepts and the non-ICN state-of-the-art for reference;
section \ref{sec:enabler} presents different technologies that enable
distributed computing in ICN; section \ref{sec:protocol} analyzes
various protocols dedicated to in-network computing; section
\ref{sec:orchestration} summarizes scheduling and computing
approaches; section \ref{sec:application} presents ICN-enabled
distributed computing applications; and section \ref{sec:conclusion}
discusses lessons learned and suggests future directions for
distributed computing.

%% file: discompicn.tex
\section{Distributed Computing in ICN}
\label{sec:discompicn}

Distributed computing has different facets, for example, client-server
computing, web services, stream processing, distributed consensus
systems, and Turing-complete distributed computing platforms. There
are also different perspectives on how distributed computing should be
implemented on servers and network platforms, a research area that we
refer to as {\em Computing in the Network}.
Active Networking~\cite{tennenhouse1996towards}, one of the earliest
works on computing in the network, intended to inject programmability
and customization of data packets in the network itself; however,
security and complexity considerations proved to be major limiting
factors, preventing its wider deployment~\cite{irtf-coinrg-dir-00}.
{\em Dataplane programmability} \cite{10.1145/2656877.2656890} refers
to the ability to program behavior, including application logic, on network elements and SmartNICs, thus enabling some form in-network computing. Alternatively,
different types of server platforms and light-weight execution
environments are enabling other forms of distributing computation in
networked systems, such as architectural patterns, such as {\em edge
  computing}.

In this study, we focus on distributed computing and on how
information-centricity in the network and application layer can
support the development and operation of such systems. The rich set of
distributed computing systems in ICN suggests that ICN provides some
benefits for distributed computing that could offer advantages such as
better performance, security, and productivity when building
corresponding applications. We discuss issues in legacy distributed
computing in section \ref{sec:legacy-issues}, ICN features to address
some of these issues in section \ref{subsec:architecture}, and
performance metrics that are frequently used by different ICN
distributed computing approaches in section \ref{sec:metrics}.

\subsection{Issues in Legacy Distributed Computing}
\label{sec:legacy-issues}

Although there are many distributed computing applications, it is also
worth noting that there are many limitations and performance issues
\cite{truong2018performance}. Factors such as network latency, data
skew, checkpoint overhead, back pressure, garbage collection overhead,
and issues related to performance, memory management, and
serialization and deserialization overhead can all influence the
efficiency. Various optimization techniques can be implemented to
alleviate these issues, including memory adjustment, refining the
checkpointing process, and adopting efficient data structures and
algorithms.

Some performance problems and complexity issues stem from the overlay
nature of current systems and their way of achieving the
above-mentioned mechanisms with temporary solutions based on TCP/IP
and associated protocols such as DNS. For example, Network Service
Mesh has been characterized as architecturally complex
\cite{infoqFutureService} because of the so-called {\em sidecar}
approaches and their implementation problems.

In systems that are layered on top of HTTP or TCP (or QUIC), compute
nodes typically cannot assess the network performance directly -- only
indirectly through observed throughput and buffer
under-runs. Information-centric data-flow systems, such as IceFlow
\cite{10.1145/3460417.3482975}, claim to provide better visibility and
thus better joint optimization potential by more direct access to
data-oriented communication resources. Then, some coordination tasks
that are based on exchanging updates of shared application state can
be elegantly mapped to named data publication in a hierarchical
namespace, as the different dataset synchronization (Sync) protocols
(discussed in section \ref{sec:Dataset Synchronization}) in NDN
demonstrated. \cite{Moll2022SoKTE}


\subsection{Information-Centric Distributed Computing}
\label{subsec:architecture}

ICN generally attempts to provide a more useful service to
data-oriented applications but can also be leveraged to support
distributed computing specifically.

\paranl{Names} Accessing named data in the network as a native service
can remove the need for mapping application logic identifiers such as
function names to network and process identifiers (IP addresses, port
numbers), thus simplifying implementation and run-time operation, as
demonstrated by systems such as Named Function Networking (NFN)
\cite{Tschudin2014NamedFA}, RICE \cite{Krl2018RICERM}, and IceFlow
\cite{10.1145/3460417.3482975}. It is worth noting that, although ICN
does not generally require an explicit mapping of names to other
domain identifiers, such networks require suitable forwarding state,
e.g., obtained from configuration, dynamic learning, or routing.

\paranl{Data-orientedness}
ICN's notion of immutable data with strong name-content binding
through cryptographic signatures and hashes seems to be conducive to
many distributed computing scenarios, as both static data objects and
dynamic computation results in those systems such as input parameters
and result values can be directly sent as ICN data objects. NFN has
first demonstrated this.

Securing distributed computing could be supported better in so far as ICN does not require additional dependencies on public-key or pipe
securing infrastructure, as keys and certificates are simply named
data objects and centralized trust anchors are not necessarily needed
\cite{Fu2018InformationCentricNS, AbdAllah2015ASO}. 
Larger data collections can be aggregated and re-purposed by manifests
(FLIC, \cite{irtf-icnrg-flic-04}), enabling ``small'' and ``big data''
computing in one single framework that is congruent to the
packet-level communication in a network. IceFlow uses such an
aggregation approach to share identical stream processing results
objects in multiple consumer contexts.

Data-orientedness eliminates the need for connections; even reliable
communication in ICN is completely data-oriented. If higher-layer
(distributed computing) transactions can be mapped to the network layer
data retrieval, then server complexity can be reduced (no need to
maintain several connections), and consumers get direct visibility
into network performance. This can enable performance optimizations,
such as linking network and computing flow control loops (one
realization of {\em joint optimization}), as showed
by IceFlow.

\paranl{Location independence and data sharing} Embracing the
principle of accessing named and authenticated data also enables
location independence, i.e., corresponding data can be obtained from
any place in the network, such as replication points ({\em repos}) and
caches. This fundamentally enables better multi-source/path
capabilities as well as data sharing, i.e., multiple data retrieval
operations for one named data object by different consumers can
potentially be completed by a cache, repo, or peer in the network.

\paranl{Stateful Forwarding} ICN provides stateful, symmetric
forwarding, which enables general performance optimizations such as
in-network retransmissions, more control over multipath forwarding,
and load balancing. This concept could be extended to support
distributed computing specifically, for example, if load balancing is
performed based on RTT observations for idempotent remote-method
invocations.

\paranl{More Networking, less Management} The combination of
data-oriented, connection-less operation, and stateful (more powerful)
forwarding in ICN shifts functionality from management and
orchestration layers (back) to the network layer, which can enable
complexity reduction, which can be especially pronounced in
distributed computing. For example, legacy stream processing and
service mesh platforms typically must manage connectivity between
deployment units (pods in Kubernetes \cite{10.1145/3539606}). In
Apache Flink \cite{carbone2015apache}, a central orchestrator manages
the connections between {\em task managers} (node agents). Systems
such as IceFlow have demonstrated a more self-organized and
decentralized stream-processing approach, and the presented principles
are applicable to other forms of distributed computing.

\paranl{Summary} In summary, we can observe that ICN's general
approach of having the network providing a more natural (data
retrieval) platform for applications benefits distributed computing in
similar ways as it benefits other applications. One particularly
promising approach is the elimination of layer barriers, which enables
certain optimizations. In addition to NFN, there are other approaches
that jointly optimize the utilization of network and computing
resources to provide network service mesh-like platforms, such as edge
intelligence using federated learning \cite{Rahman2022OnTI}, advanced
CDNs where nodes can dynamically adapt to user demands according to
content popularity \cite{Ghasemi2020iCDNAN, Narayanan2018OpenCDNAI},
and general computing systems \cite{Krl2019ComputeFN,
  Mastorakis2020ICedgeWE, Kutscher2021VisionID}.

\subsection{System Metrics}
\label{sec:metrics}

\input{tables/metrics.tex}

Information-centric distributed computing systems can provide a range
of optimizations, depending on the type of applications they support
and the environment in which they are intended.  Systems built on top
of edge environments, such as ICedge \cite{Mastorakis2020ICedgeWE},
focus on low latency for distributed computing by leveraging knowledge
about spatial proximity in edge computing scenarios, addressing
applications such as IoT, mobile computing, Extended Reality (XR), and
vehicular distributed computing that can benefit from offloading
computation from cloud-based servers to the edge. A brief overview of
these metrics is presented in Table \ref{tab:metrics}.  Latency is a
key metric in information-centric distributed-computing systems. The
end-to-end delay is mainly composed of two parts: 1) the time consumed
by data packet transmission in the network, and 2) the execution time
of computing tasks. In addition to studies dedicated to network
transmission time \cite{Wang2016C3POCC, Ullah2020ICNWE} or computation
completion time \cite{Fan2021ServingAT, Krl2019ComputeFN}, most
studies measure the total latency (end-to-end delay) \cite{Qi2021R2AD,
  Krl2017NFaaSNF, Tan2019SchedulingOD, Kanai2022InformationcentricSM,
  Mastorakis2020ICedgeWE, Sifalakis2014AnIC, Ambalavanan2022DICerDC}
in their evaluations. It is worth noting that only a few works
\cite{Sifalakis2014AnIC, Yamamoto2019MultipleNF, ibn2017optimal} have
evaluated the network and computation latency respectively.
To better understand systems and their optimization potential, as well
as for run-time optimization, it is important to analyze network and
computation latency separately.  Fine-granular latency metrics can
enable more accurate resource control and help us understand which
resource optimization and mechanism (caching, load balancing, etc.)
contributes to performance improvements. Section \ref{sec:programming}
discusses this from the perspective of programming models.

It is also worth noting that the distributed system should not only
limit the optimization goal of the system to performance (such as
latency, overhead, and balance), but also pay more attention to other
metrics, especially the native security and robustness provided by
ICN. Although latency accounts for a large proportion of the
objective, other metrics are also being explored. For example,
\cite{Fan2021ServingAT, Mastorakis2020ICedgeWE} considered overhead in
terms of the number of sent packets (\textit{Interest} and responses) and
forwarder state size as important indicators for evaluating their
systems. Load balancing is a key technology for improving the
performance; NFaaS \cite{Krl2017NFaaSNF} and C3PO
\cite{Wang2016C3POCC} are concerned with compute load
balancing. Moreover, there are other perspectives, including fault
tolerance \cite{Scherb2017ExecutionSM, Krl2019ComputeFN,
  Mastorakis2020ICedgeWE}, satisfaction rate \cite{Krl2017NFaaSNF},
delay evolution \cite{Krl2017NFaaSNF}, drop rate
\cite{Wang2016C3POCC}, times of computation result reuse
\cite{Tan2019SchedulingOD}, and resource utilization
\cite{Ambalavanan2022DICerDC}.

%% file: tables/metrics.tex
\begin{table}[htpt]
  \caption{Metrics Overview}
  \label{tab:metrics}
  \resizebox{\columnwidth}{!}{%
  \begin{tabular}{lll}
  \hline
  \textbf{Metrics}     & \textbf{Related Works}                                                                                                                                                                                                                                                        & Count \\ \hline
  End-to-end Latency   & \begin{tabular}[c]{@{}l@{}}R2 \cite{Qi2021R2AD}, NFaaS \cite{Krl2017NFaaSNF},  \cite{Tan2019SchedulingOD}, \\ \cite{Kanai2022InformationcentricSM},  ICedge \cite{Mastorakis2020ICedgeWE},  NFN \cite{Sifalakis2014AnIC}, \\ DICer \cite{Ambalavanan2022DICerDC}\end{tabular} & 7     \\ \hline
  Fault Tolerance      & \cite{Scherb2017ExecutionSM}, ICedge \cite{Mastorakis2020ICedgeWE}, CFN \cite{Krl2019ComputeFN}                                                                                                                                                                               & 3     \\ \hline
  Transmission Latency & \begin{tabular}[c]{@{}l@{}}C3PO \cite{Wang2016C3POCC}, \\ ICN with edge for 5G \cite{Ullah2020ICNWE}\end{tabular}                                                                                                                                                             & 2     \\ \hline
  Computing Latency    & Serving at the Edge \cite{Fan2021ServingAT}, CFN \cite{Krl2019ComputeFN}                                                                                                                                                                                                      & 2     \\ \hline
  System Overhead      & \begin{tabular}[c]{@{}l@{}}Serving at the Edge \cite{Fan2021ServingAT}, \\ ICedge \cite{Mastorakis2020ICedgeWE}\end{tabular}                                                                                                                                                  & 2     \\ \hline
  Load Balance         & NFaaS \cite{Krl2017NFaaSNF},  C3PO \cite{Wang2016C3POCC}                                                                                                                                                                                                                      & 2     \\ \hline
  Latency Evolution    & NFaaS \cite{Krl2017NFaaSNF}                                                                                                                                                                                                                                                   & 1     \\ \hline
  Resource Utilization & DICer \cite{Ambalavanan2022DICerDC}                                                                                                                                                                                                                                           & 1     \\ \hline
  Satisfication Rate   & NFaaS \cite{Krl2017NFaaSNF}                                                                                                                                                                                                                                                   & 1     \\ \hline
  Compute Result Reuse & \cite{Tan2019SchedulingOD}                                                                                                                                                                                                                                                    & 1     \\ \hline
  Drop Rate            & C3PO \cite{Wang2016C3POCC}                                                                                                                                                                                                                                                    & 1     \\ \hline
  \end{tabular}%
  }
  \end{table}

%% file: enabler.tex
\section{Enablers}
\label{sec:enabler}

ICN distributed computing systems are using several approaches and
technologies. In this section, we discuss naming conventions,
computing and performance, security and trust, and availability and
tolerance.

\subsection{Naming Conventions}
\label{sec:naming}

\begin{figure}[htbp]
    \centering
    \includegraphics[width=0.4\textwidth]{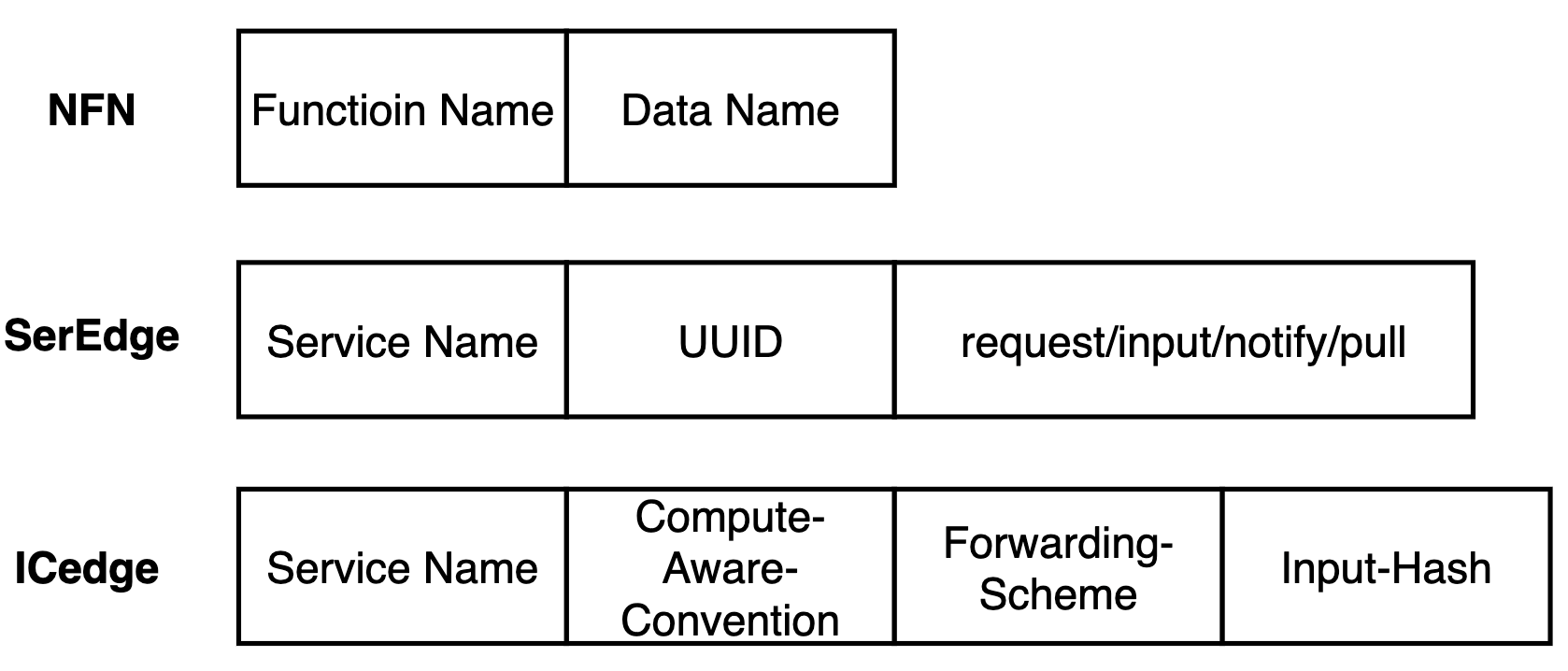}
    \caption{Three Examples of Naming Conventions, including lambda in NFN \cite{Sifalakis2014AnIC} hierarchical in Serving at the Edge \cite{Fan2021ServingAT}, flat in ICedge \cite{Mastorakis2020ICedgeWE}.}
    \label{fig:naming}
\end{figure}

Naming a computing \textit{Interest} or process is a key concept in NFN or
general distributed computing systems over ICN. As shown in figure
\ref{fig:naming}, a straightforward method is to encode a lambda
function into the original data name \cite{Sifalakis2014AnIC}.
However, this simple naming convention cannot cater to complex
computing process combinations and their input parameters. Many
systems have proposed their own naming conventions to support complex
or durable computing \textit{Interests}.
Many distributed systems name their object and computations in a
hierarchical way for easier service discovery and reuse of computation
results. For example, \cite{Fan2021ServingAT} proposes constructing
names by concatenating service identifier, session identifier, and
operating information. This convention is suitable for service
\textit{Interest} forwarding because it decouples services from server
nodes. ICedge \cite{Mastorakis2020ICedgeWE} uses namespaces to perform
service discovery. This can facilitate service mesh implementation and
a seamless user onboarding experience. More details are available in
these surveys on naming conventions: \cite{Tehrani2022SoKPK,
  Serhane2021ASO}.

\subsection{Programming Models}
\label{sec:programming}

Functional programming abstracts computation as the process of
mathematical function evaluation. The idea that functions are
first-class citizens corresponds well with ICN's data-oriented
architecture, in which data is the first-class citizen. Therefore,
computation tasks are offloaded in a modular and declarative
manner. Lambda is a function that can be encoded into ICN names
\cite{Sifalakis2014AnIC, Tschudin2014NamedFA}. Although functional
programming provides a powerful computing abstraction, other issues
exist, such as simplifying concurrency management. Distributed
computing systems face a similar challenge: How can multiple nodes be
scheduled to work on the same task at the same time collaboratively?
The actor model and message-oriented communication, such as in
SmallTalk \cite{Goldberg1983Smalltalk80TL} and Erlang
\cite{Armstrong1993ProgrammingES}, try to structure systems and their
communication processes to support this. In ICN, IceFlow
\cite{Kutscher2021VisionID} uses the actor model and dataflow concepts
for parallel processing.

In the actor model, independent computing entities communicate with
one another through messages. Actors own states and behaviors and can
create other actors. This model enables high degrees of concurrency,
scalability, and fault tolerance. Dataflow represents the task as a
directed graph where nodes and edges represent computations and data
flows, respectively. At the logical layer, dataflow systems typically
provide multi-destination communication, as the results can be
consumed by many downstream actors. This matches well with ICN's
data-oriented operation and in-network replication and can be further
supported by ICN caching. On the other hand, ICN's pull-based
communication is not conducive to fundamentally push-based dataflow
communication.  IceFlow \cite{Kutscher2021VisionID} addresses this by
employing Sync approaches based on periodic
pulling. We suggest that some push-based communication mechanisms can
be adopted by dataflow-based distributed computing systems, which we
discuss further in section \ref{subsec:event-driven}.

Computation re-use is important for improving distributed computing systems to reduce computing redundancy. \cite{Mastorakis2020ICedgeWE} defines a set of compute-aware naming conventions that cluster the computation task \textit{Interests} at compute nodes: Subsequent identical computation requests can be forwarded to the specific compute nodes where the results already exist. \cite{Tan2019SchedulingOD} also aggregates identical service requests and evaluates the number of result reuse. In general, the ICN network layer can provide general caching, but cannot assist with computation re-use, which requires some computing system knowledge.  Current ICN systems generally lack the ability to assess the computation re-use potential, which calls for the development of naming conventions for expressing this potential explicitly, as formulated by RICE \cite{Krl2018RICERM}.

\subsection{Security and Trust}
\label{subsec:security}

Security and trust are important enablers of distributed computing
systems, especially in multiple tenant environments.  Inherited from
ICN, data-oriented security in distributed computing offers precise
access control, confidentiality, and integrity protection at the data
level by binding names and contents with digital signatures. This
enables secure data sharing and resilience to system changes, thereby
enhancing the overall security and reliability of distributed
computing systems. Building on this, some works start from
metadata. \cite{Liu2016DemonstrationOA} proposes the use of signatures
and hash codes to guarantee the security of the results from a
functional chaining system. 
\cite{Marxer2016AccessControlledIP} adopts an access control list (ACL) structure to list all permited client identities. Through content production chains where results are produced out of results, a named object will be modified on the content load part but ACL. That means ACL keeps unchangeble to remark permited identities.
Some studies have started at the protocol level. For example, RICE
\cite{Krl2018RICERM} designed a two-stage handshake protocol to
provide consumer authentication and authorization.

In general, if the results of remote execution tasks are recalculated
locally to verify credibility, this violates the original intention of
sending the request. \cite{Marxer2020ResultPI} argues that the
validation of data is about verifying authenticity, while the
resolution result is about correctness. This decouples authenticity
from correctness, assuming that all participants are honest and that
the results should be correct. This concept achieves the first
verification mechanism results without recomputing the task. A
decrease in the cost of verification will promote the use of
distributed computing systems.

\subsection{Availability and Tolerance}
\label{subsec:availability}

A distributed computing system usually comprises heterogeneous
resources, such as different architectures (ARM, X86, etc.) and
functions (routers, servers, etc.). In addition, edge computing
environments are usually physically fragile and unreliable, for
example considering edge network mobility, battery exhaustion, and
link failure. These factors make the availability of distributed
computing systems, for example in edge networks, a key issue. Several
studies have focused on this issue. \cite{Scherb2017ExecutionSM}
provides a straightforward solution to interact with the Process
Control Block (PCB) of the request computing process to achieve fault
detection, task cancelation, task stopping, and task
resuming. However, this solution only tackles state detection and
modification and does not provide architecture-level support.

To resolve this, CFN \cite{Krl2019ComputeFN} has been designed to
tolerate computing node failure and the corresponding loss of \textit{Interest}
response. The framework can either choose to recompute the request
upon failure detection in a proactive manner or defer the retry until
the result is eventually required. The latter strategy is similar to
function resolution in functional programming languages. This approach
increases the complexity and the overhead of the system. Mastorakis et
al.\cite{Mastorakis2020ICedgeWE} claimed that avoiding a single point
of failure requires a backup mechanism, and this method increases the
synchronization overhead. They moved the fault tolerance function from
the overlay system to the underlay network, which can implement
forwarding-based recovery to avoid node failure. Our analysis
indicates that fault tolerance should consider not only the task level
but also both the node and network levels. Node-level solutions, such
as those in CFN \cite{Krl2019ComputeFN}, follow the idea of
traditional distributed systems. Because CFN is built on top of ICN, a
flexible forwarding plane is a reliable yet simple solution to achieve
this goal.

%% file: protocol.tex
\section{Protocols}
\label{sec:protocol}

This section describes the protocols for communication between
distributed processes in ICN. Generally, these protocols enable
applications to leverage computing and communication resources
distributed in the network via higher-level APIs. We also consider the
role of metadata and data structure in ICN protocols for distributed
content delivery.

\subsection{Remote Method Invocation}
\begin{figure*}
    \centering
    \includegraphics[width=\linewidth]{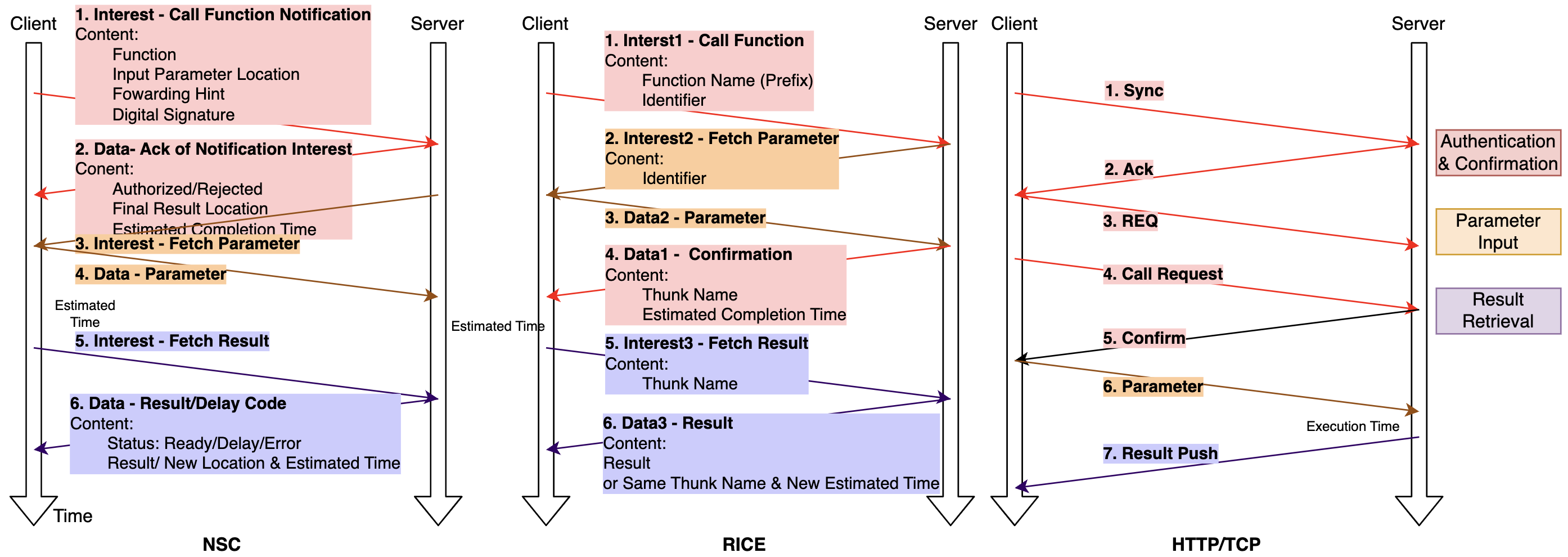}
    \caption{A Brief Comparision of NSC \cite{meirovitch2021nsc}, RICE \cite{Krl2018RICERM}, and HTTP/TCP RMI}
    \label{fig:rmis}
\end{figure*}

RMI is enabling client-server and
peer-to-peer distributed computing scenarios, typically with the goal
to provide a transparent funcation invocation service, where the
invoking application does not have to be aware of the distributed
nature of the system. In this section, we introduce several ICN RMI
protocols. One demo by Yamamoto et al. \cite{Yamamoto2019MultipleNF}
designed a protocol capable of performing the same function on
multiple IoT devices such as finding objects on multiple smart
cameras. Similarly, DNMP \cite{Nichols2019LessonsLB} defines a Pub/Sub
protocol to send measurement functions to multiple devices for
execution and to retrieve execution results, based on a series of
libraries facilitating distributed measurement in NDN.
In Named Service Calls (NSC) \cite{meirovitch2021nsc} (figure
\ref{fig:rmis}-NSC), a client will publish its RMI interest as data
and subscribe to the corresponding result name. An NDN Sync protocol,
syncps, was adopted to update all servers with information about newly
published RMI \textit{Interests}. Subsequently, all subscribed servers execute
the function and publish the result to the name to which the client
has subscribed. Finally, the client retrieves the result. Unlike DNMP,
RICE \cite{Krl2018RICERM}, as shown in figure \ref{fig:rmis}-RICE, NSC
regards RMI requests as \textit{Interests} rather than data.  A four-way
handshake protocol is used to establish an interactive session-like
connection, and the paired clients and servers perform subsequent
RMIs. RICE inserts a second, reverse, \textit{Interest}-data interaction (for
parameter transfer) between the first interaction (for authentication
and RMI confirmation). In contrast to RICE, NSC
\cite{meirovitch2021nsc}, as shown in figure \ref{fig:rmis}-NSC,
adopts NDN's native security features to authenticate the client and
server to provide a drop-in framework for many scenarios. NSC and RICE
have similar parameter input and result retrieval processes. They both
request parameters from the server and retrieve results using a {\em
  thunk} or result location after an estimated execution time. As
shown by the arrow directions in figure \ref{fig:rmis}, NSC has
bidirectional communication between the client and server, whereas
RICE has unidirectional communication from the client to the
server. This means that NSC requires the client to own a routable
name, whereas RICE does not. RICE maintains the anonymity of users
through pull-based operations. From the perspective of parameter input
(orange color parts), RICE inserts the parameter exchange period in
the first RMI call which needs to extend the PIT expiration time. The
extra modifications to forwarders may cause additional system
complexity.  NSC does not have this problem because it uses a separate
\textit{Interest}-data interaction for parameter transmission.

RICE further highlights the inadequacy of the basic \textit{Interest}/data
exchange model of CCNx/NDN-style ICN, specifically in the context of
RESTful communication that adheres to the principles of the REST
architectural style. This inadequacy is particularly pronounced
in scenarios involving the transmission of resource representations or
request parameters from clients to servers. RESTful-ICN
\cite{Kutscher2022RESTfulIN} and \cite{Gndoan2020TowardAR} envision an
ICN-based protocol framework that can leverage key properties of the
REST architectural pattern, such as security, consumer anonymity, and
session continuation. We see the potential of RESTful for ICN in
establishing an ICN-native infrastructure for applications,
encompassing an Information-Centric web and RMI rather than merely
charting existing HTTP mechanisms onto ICN.

\subsection{Event-Driven Communication} 
\label{subsec:event-driven}

Event-driven protocols facilitate the exchange of information between
different endpoints based on asynchronous events (that could trigger
computation), instead of continuous or periodic communication as
observed in request-reply protocols. ICN has a two-fold relationship
to even-driven communication: On the one hand side, asynchronously
generated data could be treated as named data as well, so the
fundamental principle of accessing named data in the network seems
conducive to event-driven communication, especially when considering
that CCNx/NDN-based architectures feature intrinsic data replication
and \textit{Interest} aggregation (which might resemble Pub/Sub
application-layer systems such as MQTT). On the other hand,
communication in CCNx/NDN-based ICN is fundamentally {\em pull-based},
i.e., based on \textit{Interest}/data interactions, which is the opposite of
event-driven communication. We can distinguish three lines of work in ICN:

1) {\bf Information-Centric architectures that are itself based on the
  Pub/Sub paradigm} which could provide more natural support for
using this paradigm for distributed computing:
DONA~\cite{10.1145/1282427.1282402} and
PSIRP~\cite{DBLP:conf/fia/TarkomaAV09} are examples of such
architectures where distributed computing is leveraged on the
infrastructure layer: locating objects with topology-independent
identifiers across heterogeneous inter-domain settings (i.e. different
administrative domains) and inter-domain rendezvous services that
combine policy-based name routing between adjacent networks with
hierarchical interconnection overlays for locating objects across
different domains~\cite{rajahalme2011name}.

2) {\bf Pub/Sub extensions to CCNx/NDN-based systems} to ameliorate
the above-mentioned incongruence of pull-based protocols and push
semantics: Content-based Publish/Subscribe Networking
\cite{carzaniga2011content} combines pull (on-demand) with
pub/sub-based communication by extending forwarders and ICN protocols,
so that producers can register prefixes for which they would generate
data asynchronously in the network. COPSS~\cite{6062723} provides a
scalable pub/sub service using multicast concepts. At the ICN
forwarding layer, COPSS uses a multiple-sender, multiple-receiver
multicast capability, similar to PIM-Sparse-Mode, relying on
Rendezvous Points (RP). Users subscribe to content based on so-called
Content Descriptors (CDs). COPPS-aware routers provide additional data
structures, i.e., a subscription table for maintaining CD
subscriptions. Another approach that provides support for both push
and pull in ICN is HoPP~\cite{gundougan2018hopp}, which adopts a
hybrid architecture consisting of brokers and peers. Brokers
communicate in a centralized manner to manage subscriptions, and peers
communicate with each other in a decentralized manner to exchange
data, while gateways bridge the gap between the two types of
components. This hybrid architecture balances scalability and
efficiency.

3) {\bf ICN Distributed Computing systems leveraging Pub/Sub}
NFN~\cite{scherb2017execution} provides a mechanism to control
long-lasting in-network computations in ICN by enabling debugging,
timeout prevention (i.e. changing timeout on the fly), fetching
intermediate results and client-side computation steering. NFN uses
pub-sub to fetch intermediate computing results. IceFlow~\cite{Kutscher2021VisionID} leverages dataset
synchronization (Sync) to inform dataflow nodes about newly available
input data, which then triggers new computations.

Unlike CCNx/NDN-based ICN, DONA, and PRIRP, which only enable either
pull-based or push-based interactions,
INetCEP~\cite{DBLP:conf/ancs/LuthraKHLRKF19} enables both interaction patterns in a single ICN architecture via an expressive complex event processing (CEP) query language and a CEP query engine. CEP query language can distinguish between pull and push-based trafﬁc and supports standard
operators such as windows, joins, filters and aggregators. The CEP
query engine can execute these operators on CEP queries in both a
centralized and distributed fashion.

\subsection{Dataset Synchronization}
\label{sec:Dataset Synchronization}

Multiparty communication and coordination are essential for a
distributed system or application, and ICN generally supports this
well. For example, ICN uses data object names for de-multiplexing and
enables receivers to fetch data according to their individual needs
and capabilities. However, participants must know the
collection of data item names in advance.
Sync is a
commonly used transport service for reliable name synchronization in
some distributed computing approaches. Sync enables dataset
synchronization through updates to a common (shared) namespace, and
different variants provide different mechanisms for efficient
pull-based update protocols.
For example, SVS \cite{Li2018ABI}, as one vector-based sync protocol,
encodes the dataset state in state vectors where a vector represents a
participant. Participants perform synchronization by event-driven or
periodic dataset state exchange.
Sync introduces a trade-off between delay and overhead. More frequent
synchronization operations (pull) can reduce latency but also increase
the overhead. Furthermore, current Sync protocols operate in a
decentralized manner, leveraging \textit{Interest} multicast forwarding, which
can pose implementation and deployment challenges in non-local
networks. A detailed discussion can be found in \cite{Moll2022SoKTE}
and \cite{Shang2017ASO}.

\subsection{Metadata and Data Structures}
\label{sec:metadata}

Well-designed metadata and data structures can help improve the
computational efficiency and reduce latency. For example, the R2
remote function execution mechanism~\cite{Qi2021R2AD} attempts to
select the best executor in a distributed manner. Metadata is used to
describe the input data which consumes network resources. R2 uses a
cost estimation model for jointly optimizing network and compute
resources. During the process of pulling metadata, the system can also
``warm up'' the execution environment. Therefore, the optimal node
selection and hot runtime provisioning in R2 reduce the end-to-end
computation completion time. CFN \cite{Krl2019ComputeFN} adopts a
compute graph to represent the computation process, which helps
coordinate distributed computing nodes to work collaboratively on the
same task. Similarly, Tangle-Centric Networking
(TCN)~\cite{Scherb2021TangleCN} proposes a decentralized data
structure for coordinating distributed applications. In essence, these
metadata or data structures are abstractions for information sharing
among ICN networks that can help to support decision making in
distributed computing.

%% file: orchestration.tex
\section{Orchestration}
\label{sec:orchestration}

In distributed computing systems, orchestration automates and
coordinates various components across nodes to achieve common
objectives. Its primary tasks include resource allocation and
provisioning, task scheduling and scaling, configuration management,
service deployment, and state monitoring.

\subsection{Scheduling}
\label{subsec:schedule}

In distributed computing, resources are partitioned and allocated to
different applications and services. Task scheduling is the process of
allocating resources to tasks to achieve the performance objectives as
described in section \ref{sec:metrics}. In this section, we
discuss provisioning and scheduling approaches in ICN-based
distributed computing.

\subsubsection{Resource Allocation and Provisioning}

Fine-grained resource provisioning approaches are more flexible and
responsive than coarse-grained ones; although they also introduce more
challenges into the system, such as resource management
complexity. Leveraging ICN's name-based operation, resource
provisioning and allocation can be performed in a more flexible and
fine-grained manner. Here, we present a brief overview of granularity
systems built on ICN. From coarse to fine granular,
\cite{Tan2019SchedulingOD} uses Raspberry Pis, i.e., physical
machines, to provide a runtime environment for NDN's Forwarding Daemon
(NFD). \cite{Ullah2020ICNWE} uses a virtual machine (VM) as the
runtime of ndnSim \cite{Mastorakis2017OnTE}. \cite{Fan2021ServingAT,
  Kanai2022InformationcentricSM} use Docker containers to provide
services. \cite{Krl2017NFaaSNF} adopts unikernels
\cite{Madhavapeddy2013UnikernelsRO}, a lightweight virtualization
technology compared to containers and VMs
\cite{Plauth2017APE}. \cite{Krl2019ComputeFN, Sifalakis2014AnIC}
provide computing resources through worker processes while
\cite{Kutscher2021VisionID, Krl2019ComputeFN} both use actors as the
basic resource unit. We argue that systems that are based on ICN and
thus have the ability to distribute fine-grained tasks should adopt
resource allocation methods that can match their granularity. For
example, creating a VM for a function request is not worth the cost,
while performing tasks in VMs with a persistent context in traditional
distributed computing will be more efficient.

\subsubsection{Service Deployment}

Using resource-provisioning technologies, schedulers (centralized or
decentralized) can optimize where and when to place services or
executors. For example, \cite{Ullah2020ICNWE} allocates resources and
prefetch caches in radio network stations according to the popularity
of content and services. \cite{ibn2017optimal} proposes an optimal
placement algorithm for Hadoop over ICN to minimize net data transfer
and computation costs. CFN \cite{Krl2019ComputeFN} divides resources
into several pools of compute nodes. A scoped flooding resource
advertisements protocol was designed for dynamic computation serving
each node. \cite{Fan2021ServingAT} modeled the popularity of edge
computing services and designed a service dynamic pull strategy. This
mechanism instantiates services to minimize the request completion
time. Similarly, NFaaS \cite{Krl2017NFaaSNF} uses a pull strategy to
pull service unikernels to nodes where this service is popular. These
methods consider the dynamicity of service popularity and adapt to
changes, which leads to a responsive system.

\subsection{Configuration}

IP-based systems must design complicated configuration and scalability
components to address and service resolution as well as scalability
issues. ICN-based methods without locators can avoid some of these
problems. IceFlow \cite{Kutscher2021VisionID} uses Sync for managing
dataflow components in a decentralized manner. Unlike conventional
dataflow systems, this approach operates without complex
configuration, leveraging distributed data structures and
synchronization protocols, as described in section
\ref{sec:metadata}. Configuration is often considered an engineering
problem, and is thus neglected in research. However, the ease of use
of a system determines the utility of a technology to a large extent,
and its ease of use is clearly dependent on flexible and simple
configuration mechanisms.

\subsection{Compute Mesh}
\label{sec:compute-mesh}

{\em Service mesh} is a dedicated service composition approach that is
based on service-to-service communications, including service
discovery and dispatching. Given the similarity between service and
compute functions in the context of distributed computing systems, we
define \textit{Compute Mesh} as a more general concept of {\em service mesh}.
With respect to computing discovery, requesters in ICedge
\cite{Mastorakis2020ICedgeWE} send \textit{Interests} under namespace
$/discovery$, and then compute nodes send a response back to the
requesters with metadata describing how to call the services. We
observe that most systems adopt a namespace approach to perform
service discovery. This match with ICN can reduce the reliance on more
complex middleware in conventional systems.

Regarding computation offloading and best executor selection as a
multi-objective problem, approaches such as
\cite{Ambalavanan2022DICerDC, Mastorakis2020ICedgeWE,
  Fan2021ServingAT} all model their objective as cost and design
different strategies, such as proactive and passive
\cite{Mastorakis2020ICedgeWE}, to forward the computing request to an
optimal node executor. For example, \cite{Ambalavanan2022DICerDC}
designed a distributed coordination mechanism to achieve computation
resolution with respect to resource utilization rate and computation
completion time. Another important observation is that most approaches
adopt a synchronization data structure to update information among
nodes, such as SVS \cite{Li2018ABI} in DICer
\cite{Ambalavanan2022DICerDC}, compute graphs, and PSync
\cite{Zhang2010NamedDN} in \cite{Krl2019ComputeFN}. However, flooding
in scoped resource pools has also been adopted for resource
advertisement and update \cite{Krl2019ComputeFN}. We can observe that
ICN can support {\em compute mesh} well because both naming
conventions, as described in section \ref{sec:naming}, and name-based
routing can be leveraged to elegantly support specific objectives.

%% file: application.tex
\section{Applications}
\label{sec:application}

Various ICN distributed computing applications have been developed. We
categorize them and describe their properties with respect to how they
leverage information-centricity.

\subsection{Distributed Microservices Communication } 

Microservices offer a paradigm for constructing complex software applications by dividing them into small, independent, and loosely coupled units. However, as applications scale and the number of microservices multiply, intricate inter-service communication becomes a challenge. Traditional centralized service-oriented network architectures often fail to handle the complexities of extensive distributed microservices deployments. This inadequacy manifests in multiple dimensions: performance bottlenecks, owing to increased latency when routing traffic through a central controller; scalability issues, as expanding a centralized control point becomes increasingly complex and requires substantial resources and meticulous management; and reliability constraints, where any downtime or malfunction in the central control component can disrupt the entire inter-service communication flow \cite{li-icnrg-damc-01}.

Some approaches have made attempts at simplifying microservice systems by leveraging ICN's name-based, locator-less operation. µNDN \cite{marchal2018mu} implements a management plane to monitor and orchestrate microservices according to pre-defined rules. This system disaggregates the key functions of an NDN forwarder into specialized Virtualized Network Functions (VNFs) and utilizes the VNF manager for adaptive deployment. Although effective, this approach adds a layer of management complexity, necessitating increased orchestration and chaining tasks to maintain the network. DAMC \cite{li-icnrg-damc-01} proposes leveraging name-based operation with unique service prefixes to enhance microservice communication. It introduces multiple architectural entities, such as Service Gateway (SG), Service Router (SR), Service Prefixes Authentication (SPA), and Service Mesh Communication Scheduling Center (SCSC), and coordinates communication through well-defined control signaling messages and functions, providing a powerful framework for managing the complex communication requirements of distributed microservices. However, challenges remain \cite{imran2023mia,long2017icn,marchal2018mu}. For instance, scalability issues might arise in scenarios with uneven service popularity or applications burgeoning in size.

\subsection{Decentralized Blockchain Transactions}

Blockchain's core data dissemination mechanism—gossip protocol over P2P—often leads to efficiency deficits \cite{thai2022design}: 1) redundant data transmission over multiple TCP connections, leading to unnecessary network traffic; and 2) suboptimal routing due to nodes' lack of awareness of the physical network topology. While the first issue is manageable for small transactions, it becomes problematic for larger-block deliveries. Modifying the block propagation protocol can mitigate this, but at a cost: certain block deliveries may require 1.5 round-trip message exchanges. Additionally, because receiving peers are unaware of the nearest block source, they default to request the block from the first announcer, resulting in suboptimal delivery latency.

Implementing blockchain with ICN can leverage its in-network caching and implicit multi-destination delivery, leading to a more efficient blockchain system with enhanced data dissemination. Researchers have proposed solutions in various domains, including public key infrastructure \cite{lou2018blockchain}, data security and access control \cite{lyu2020sbac}, and vehicle networks \cite{ortega2018trusted}. Thai et al. \cite{thai2022design} proposed a protocol that utilizes a P2P overlay for data announcements and then leverages NDN's pull mechanism for retrieval, based on unique naming conventions inferred from the announcement. This unique naming of data enables request aggregation, allowing the returning data packet to be cached and replicated by forwarders in the network for efficient delivery to all requesters. Feng et al. \cite{feng2022blockchain} addressed the scalability of blockchain storage by proposing an
ICN-based approach that includes a resolution system for community division, fostering efficient blockchain node partitioning. It provides a virtual chain for faster blockchain indexing, coupled with collaborative block replica deletion, optimized for neighboring partitions. However, challenges such as specific transaction retrieval \cite{thai2022design} persist. If a piece of data is absent, the retrieval process becomes incredibly burdensome, leading to escalated transmission costs and deteriorating user experience (refer to \cite{asaf2020blockchain} for more details).

\subsection{Distributed Data Management}
 
Distributed Data Management (DDM) in NDN applies information-centric concepts and a collection of strategies to enable the manipulation, distribution, and protection of extensive geographically dispersed data \cite{wu2022n}. These strategies are employed in a variety of applications, including federated catalog systems \cite{fan2015managing}, scalable data dissemination \cite{zhang2019dledger}, version control \cite{ma2021gitsync}, and federated data repositories \cite{presley2022hydra}, enabling the effective management of large-scale, dispersed data. For example, \cite{fan2015managing} provides a federated catalog by storing and managing NDN names to accelerate the discovery of desired data across multiple domains. DLedger \cite{zhang2019dledger} ensures effective data dissemination over NDN by offering content multicast and a gating function called Proof-of-Authentication (PoA), which digitally signs records, addressing security concerns and enabling participation of constrained devices. GitSync \cite{ma2021gitsync} enables effective distributed version control through direct peer-to-peer Git synchronization. It maintains a local repository copy and runs a sync protocol, similar to SVS, in the background. This protocol broadcasts synchronization \textit{Interests}, carrying the local storage root hash, which allows connected peers to detect changes and synchronize as needed. Hydra \cite{presley2022hydra} is an NDN-based federated file storage system that enables effective federated data repositories. In Hydra, multiple file servers process and store file segments while computing a synchronously updated 'global view' of the metadata.

However, contrary to centralized TCP/IP models, some NDN-based DDM systems might incur higher synchronization costs owing to the gap between inefficient data dissemination in heterogeneous networks \cite{pan2020cdd,zhang2019dledger}, the high data throughput necessitated by P2P networks \cite{liu2021designing}, and the ICN-based pull data transmission mode (section \ref{sec:Dataset Synchronization}). Despite these challenges, NDN-based DDM systems tend to be more resilient. For instance, in Hydra \cite{presley2022hydra}, even if individual nodes are compromised, signed and encrypted files maintain their integrity and confidentiality.

\subsection{Distributed Object Stores}

In large-scale data management, Distributed Object Stores (DOS), such as Amazon S3\cite{palankar2008amazon}, Google Cloud Storage \cite{geewax2018google}, and Azure Storage\cite{soh2020azure}, are used to manage and manipulate massive data chunks across distributed infrastructure.  These DOSs organize data as unique objects that are identifiable by individual keys and spread across numerous storage servers or nodes. However, these DOSs have some limitations, primarily owing to the IP-centric locator-based communication model. A notable concern is the scaling challenge \cite{Patil2022KuaAD}, resulting from the potential need for an excessive number of TCP connections (in the worst case, N×N connections, where N is the number of storage nodes, as in the case of {\em Redis Cluster} \cite{Scaling}, a distributed implementation of Redis). In addition, there is a lack of universal IP multicast services for efficient data replication, performant transport protocols, and usable security implementations.

Chipmunk \cite{shin2020chipmunk} and Kua \cite{Patil2022KuaAD} are examples of NDN-based DOS systems. Chipmunk is an object store service over NDN that leverages effective naming schemes and unique identifiers for object-to-metadata mapping. In \textbf{Chipmunk} \cite{shin2020chipmunk}, nodes carry a common prefix, differentiated by a node ID, eliminating the need for users to know the node prefix.  Moreover, Chipmunk provides two types of storage within each node: \textit{file storage} for data, and a \textit{metadata store} for its metadata. Data segments are stored in the file system of the designated node, whereas metadata is stored in a node determined by the data identifier. \textbf{Kua} \cite{Patil2022KuaAD} approaches storage and location differently.  Kua defines the smallest storage unit as an \textit{Application Data Unit} that carries semantic meaning for the application. This allows applications to use semantically meaningful identifiers for object storage and retrieval in Kua clusters. For data location, Kua simply calculates the appropriate bucket for the data using a Forwarding Hint containing the bucket identifier. This potentially allows for better performance compared to Chipmunk owing to its simpler protocol and the absence of metadata fetching requirements.

\subsection{Content Distribution}
CDNs are overlay networks to efficiently distribute content globally by caching it near consumers for enhanced retrieval performance \cite{ghasemi2020far} and server offloading. They employ a relatively complex set of mechanisms, including request redirection and routing, which track network changes and content availability.  ICN could simplify this by leveraging some of its inherent features (e.g., in-network caching capabilities and implicit multicast) \cite{inaba2016content,ghasemi2020icdn,narayanan2018opencdn, jiang2014ncdn}.  For example, Inaba et al. \cite{inaba2016content}, employed ICN in cooperation with CDNs to enable consumers to cache contents and serve future requests for the same contents.  OpenCDN \cite{narayanan2018opencdn} implements a distributed actor-model programming approach and fosters an architecture independent of namespaces. It enables any ISP or third-party entity to collaborate in content distribution, allowing complete control over content storage and routing elements. However, these solutions rely on routing protocols to disseminate content availability information across network nodes, which poses two noteworthy issues \cite{flavel2015fastroute}. First, these protocols broadcast content availability updates to all nodes, regardless of their relevance or necessity. This can lead to potential FIB overload as the network's content volume continues to expand. Second, these protocols only calculate routes from each node to the content origins, neglecting both on-path caches, which becomes purely opportunistic, and off-path caches, whose effective utilization becomes impractical.

\subsection{Decentralized Data Sharing}

Decentralized Data Sharing (DDS) refers to the collection of applications and systems leveraging NDN to enable the distributed, peer-to-peer exchange of data. This can be applied across a range of contexts, including multimedia sharing \cite{Coomes2018AndroidMS, Gawande2019DecentralizedAS, zhang2016sharing}, augmented reality (AR) \cite{gusev2019data, burke2017browsing}, and multiserver online games \cite{moll2021quadtree, moll2019inter,wang2014demo, Chen2011GCOPSSAC}.

For example, NDNFit \cite{zhang2016sharing}, a distributed mobile health platform for DDS, showcases an exemplary approach to naming convention design. It utilizes consistent data naming with timestamps, facilitating easy \textit{Interest} construction to retrieve data for specific time intervals. With additional authentication and access control measures, this diminishes the dependence on perimeter-based security and simplifies service chaining. AR Web \cite{burke2017browsing} is an architecture that forwards signed data packets based on application-defined names. It employs media-specific coding such as layered coding \cite{nasrabadi2017adaptive} for efficient 360-degree video transmission, and "perceptual pruning" \cite{pourashraf2017perceptual} with content options represented in an NDN namespace. By offering web semantics with packet granularity, the AR web can deliver low-latency, high-granularity, and context-dependent media. Matryoshka \cite{wang2014demo} employs a bespoke naming scheme to fetch information about objects near a player and uses content caching and multicast by partitioning the virtual environment into octants. G-COPSS \cite{Chen2011GCOPSSAC} employs COPPS (section \ref{subsec:event-driven}) to enable efficient decentralized information dissemination in MMORPGs.

In the context of DDS, hierarchical names offer usability but can be limiting when it comes to representing attribute-based classification schemes typically used for organizing content. Efficient synchronization techniques (as discussed in sections \ref{sec:protocol} and \ref{sec:orchestration}) for namespace subsets would open up significant possibilities to support continuous nearest-neighbor retrieval patterns \cite{tao2002continuous} required by, for instance, AR content prefetching.

%% file: conclusion.tex
\section{Conclusions}
\label{sec:conclusion}

ICN provides an attractive platform for distributed computing. From the analyzed approaches, we can {\bf identify the following factors}: 1) accessing named data can be applied to both access to static data and dynamic computation results so that general ICN features such as name-based forwarding, locator-less operation, object security, and in-network caching can be leveraged directly. 2) ICN-empowered forwarding planes can provide additional improvements, e.g., through forwarding strategies with computing-aware algorithms and by enabling concepts such as joint resource optimization. 3) Multiparty communication, e.g., for group coordination, can be achieved without centralized servers. 4) ICN, by its security model and the other features mentioned above, is more conducive to self-managing, decentralized systems, which is also beneficial concerning complexity reduction in distributed computing.

After surveying more than 50 papers on this topic, we can conclude that the above-mentioned features often play a role in motivating different approaches. We observed the following {\bf potential for improvement} in this work: 1) Most proposed algorithms and systems are not compared with state-of-the-art systems in the non-ICN world. Admittedly, this is not always easy to do correctly (considering different software maturity levels), but we suggest more efforts be made to better understand the qualitative and quantitative potential for improvement. 2) The \textit{Interest}-Data vehicle is often used too naively for triggering computation and transmitting results (as discussed in \cite{Krl2018RICERM}). 3) Not enough real-world experiments are done, and technologies are not developed to a maturity level that would allow initial deployment and real-world experimentation.

\subsection*{Suggested Next Steps for ICN Research}

We can also derive some missing features in ICN that can better
support distributed computing.

\paranl{Asynchronous Data} Distributed computing systems often have to
deal with asynchronous events and their transmission over networks
({\em \bf push}), an interaction model that ICN does not naturally
provide. Different systems such as Sync, publish-subscribe, dataflow,
and IoT phoning-home scenarios
\cite{I-D.oran-icnrg-reflexive-forwarding} provide different
workarounds, often involving some form of active polling (unless
larger changes to the forwarding plane are proposed). In our view,
distributed computing and other applications could benefit from a
well-defined push service that works in ``point-to-point'' as well as
in group communication scenarios in Internet-scale scenarios.

\paranl{Reflexive Forwarding} RICE \cite{Krl2018RICERM} and RESTful ICN \cite{kutscher2022restful} provide an ICN-idiomatic method for client-server communication for both static data access and RMI, i.e., these systems enable longer-lasting computation and the transmission of request parameters of arbitrary size and complexity without giving up flow balance and consumer anonymity. It is achieved by a proposed ICN extension called {\em reflexive forwarding}, which involves installing temporary forwarding states on the reverse path from the producer to consumers so that the producer can request NDO from the consumer.

There is discussion on how such interactions should be done best in
ICN and how ICN-based protocols should be extended in this direction. While it is possible to construct RMI scenarios in testbed  without reflexive forwarding, ICN researchers should consider the real-world deployability concerns raised in these two papers.

\paranl{Information-Centric Web}
RESTFul ICN~\cite{kutscher2022restful} laid the foundation for an
information-centric web by enabling client/server communication with a
series of request/response interactions in a session context,
leveraging reflexive forwarding for both RESTful parameter
transmission and key exchange. This enables secure RESTful
communication using standard ICN mechanisms such as Content Object
encryption and signatures, without forcing all interactions into
TLS-like tunnels; and overall the system is supposed to provide
QUIC-like efficiency (without the connection overhead).

What is missing is the definition of, and further experimentation
with, HTTP-like protocol features, i.e., a complete RESTful protocol
that could provide a similar feature set as HTTP. Of course, an
information-centric web can go beyond HTTP's limitations. For example,
it should be explored how result parameters can be shared (for
idempotent requests) as demonstrated by RICE~\cite{Krl2018RICERM}
before.

\paranl{Scalable Dataset Synchronization} Sync in NDN
\cite{Moll2022SoKTE} is an attractive ``transport protocol service''
that can enable new ways to build distributed computing, consensus
protocols, etc. in a decentralized manner. Current Sync systems,
however, rely on \textit{Interest} multicast which has scalability and
deployability issues in the Internet. In addition, the efficiency can be low because of the update overhead, especially in larger groups with frequent namespace updates. This could be an area for future research, potentially on hybrid decentralized systems that employ relays at strategic points.

\subsection*{Distributed Machine Learning as a New Application}

Distributed ML systems could be a potential research
direction owing to the data-oriented security and communication
efficiency features offered by ICN. For example, ICN supports caching
and replication of data within the network in a decentralized
manner. This reduces not only the reliance on centralized repositories
that are vulnerable to attacks but also the centralized provider's
dominant control over the data. Furthermore, as pre-trained models
become basic building blocks for training new models or various
applications,
integrity threats \cite{Guo2022ThreatsTP} can be alleviated to an
extent by name-data signature binding security mechanisms. Regarding
network transmission efficiency, ICN can improve the efficiency of
ML systems that rely heavily on data by caching
frequently accessed datasets throughout the network. This becomes
particularly beneficial for training models that utilize common
datasets, as they can be readily available in the network once they
are cached.

Moreover, the future generation of communication system is expected to
move from Shannon’s information theory-based communication paradigm to
a semantic communication paradigm that can maximize effective
information transmission across wireless
networks~\cite{liu2023survey}. One major proposal towards this
objective is the transmission of a parameterized data model that
describes the data instead of the transmission of raw
data~\cite{chen2021distributed}. Thus, revisiting how ML is
distributed and managed is crucial for the next generation of networks
and wireless systems. We envision that ICN can play a crucial role in
the distribution of ML models for a given application in communication
networks.

%% file: ack.tex
\section*{Acknowledgments}
\label{sec:ack}

This work was supported in part by the Guangdong Provincial Key
Laboratory of Integrated Communications, Sensing and Computation for
Ubiquitous Internet of Things and by the \textit{Guangzhou Municipal
  Science and Technology Project 2023A03J0011}, \textit{Research Council of Finland, 6G Flagship program under Grant 346208}, and \textit{Business Finland Project diary number 8754/31/2022}